# Lattices from codes over $\mathbb{Z}_q$: Generalization of Constructions $D$, $D'$ and $\overline{D}$

**Eleonesio Strey** · **Sueli I. R. Costa**

**Abstract** In this paper, we extend the lattice Constructions $D$, $D'$ and $\overline{D}$ (this latter is also known as Forney's code formula) from codes over $\mathbb{F}_p$ to linear codes over $\mathbb{Z}_q$, where $q \in \mathbb{N}$. We define an operation in $\mathbb{Z}_q^n$ called zero-one addition, which coincides with the Schur product when restricted to $\mathbb{Z}_2^n$ and show that the extended Construction $\overline{D}$ produces a lattice if and only if the nested codes are closed under this addition. A generalization to the real case of the recently developed Construction $A'$ is also derived and we show that this construction produces a lattice if and only if the corresponding code over $\mathbb{Z}_q[X]/X^a$ is closed under a shifted zero-one addition. One of the motivations for this work is the recent use of $q$-ary lattices in cryptography.

## 1 Introduction

A lattice $\Lambda$ is a discrete additive subgroup of $\mathbb{R}^n$. Equivalently, $\Lambda \subseteq \mathbb{R}^n$ is a lattice if and only if there are linearly independent vectors $\boldsymbol{v}_1,...,\boldsymbol{v}_m \in \mathbb{R}^n$ such that $\Lambda$ is the set of all integer linear combinations of $\boldsymbol{v}_i$, $i = 1,...,m$, ie,

$$\Lambda = \{\alpha_1\boldsymbol{v}_1 + \cdots + \alpha_m\boldsymbol{v}_m;\ \alpha_1,...,\alpha_m \in \mathbb{Z}\}.$$

The set $\{\boldsymbol{v}_1,...,\boldsymbol{v}_m\}$ is called a base of $\Lambda$ and the number $m$ is called the rank of $\Lambda$. When $m = n$ we say that $\Lambda$ has full rank. The matrix $M$ whose rows are the vectors $\boldsymbol{v}_1,...,\boldsymbol{v}_m$

E. Strey
Department of Pure and Applied Mathematics,
Federal University of Espirito Santo, Brazil
E-mail: eleonesio.strey@ufes.br

S. I. R. Costa
Institute of Mathematics,
University of Campinas, Brazil
E-mail: sueli@ime.unicamp.br

This work was partially supported by FAPESP 2013/25997-7 and CNPq 312926/2013-8.

is called a generator matrix of $\Lambda$. The determinant of $\Lambda$ is defined as $\det\Lambda = \det(MM^t)$ and this is an invariant under change of basis. Lattices can be related to linear codes over finite rings and have been used for error correction in different contexts (see [22] and its references). We consider here $q \in \mathbb{N}$, the ring $\mathbb{Z}_q$ (integers modulo) and linear codes $C \subseteq \mathbb{Z}_q^n$, that are additive subgroups of $\mathbb{Z}_q^n$. Given a linear code $C \subseteq \mathbb{Z}_q^n$, a natural lattice $\Lambda_A(C) \subseteq \mathbb{Z}^n$ can be associated to it by considering the inverse image of the modulo $q$ mapping. $\Lambda_A(C)$ is said to be obtained from $C$ via Construction $A$ [3, 11, 15]. A lattice obtained via Construction $A$ is called a $q$-ary lattice. Those lattices have deserved special attention in recent years due to their use in the proposition of cryptographic schemes [16]. Well known problems such as the Short Integer Solution and the Ring-SIS approach in cryptography are associated to the search for the shortest vector in $q$-ary lattices [17].

Constructions $D$ and $D'$ were proposed by Barnes and Sloane in [1]. These constructions produce lattices from a nested family of binary linear codes. Lattices with good properties, such as the LDPC lattices [18] and turbo lattices [19], can be described by using the Constructions $D'$ and $D$ respectively. In the original definitions, there are restrictions on the minimum distance of the codes being used. Here, we will consider the definitions without these restrictions, because our objective is to establish these concepts in the most general possible way within our proposal.

Construction $D$ can also be associated to Construction $A$ in certain cases. In [5] it is shown that a lattice obtained via Construction $D$ from a chain of $k$ nested linear codes over $\mathbb{F}_p$ can, except for a scaling factor, always be obtained from Construction $A$ using a linear code over $\mathbb{Z}_{p^k}$.

The Construction $\overline{D}$, also called Construction by Code Formula, is a reformulation of Forney's code formula, which was introduced in [6, 7]. A set $\Gamma_{\overline{D}}$ generated by Construction $\overline{D}$ from a family of nested linear codes $C_1 \supseteq C_2 \supseteq \cdots \supseteq C_a$ over $\mathbb{F}_p$ is given by

$$\Gamma_{\overline{D}} = C_a + pC_{a-1} + \cdots + p^{a-1}C_1 + p^a\mathbb{Z}^n.$$

This construction has attracted attention since its introduction [8], [9], [13], [14], [20] and [21]. It is known that Construction $\overline{D}$ does not always produce a lattice. When $p = 2$, it has been shown recently in [13] that the Construction $\overline{D}$ produces a lattice if and only if the chain of nested linear codes is closed under Schur product. In this case, up to a scaling factor, the lattices obtained via Constructions $\overline{D}$ and $D$ are the same [13].

Construction $A'$ was proposed by Harshan et al. [8, 9]. This construction produces lattices from codes over a polynomial ring $\mathbb{F}_2[X]/X^a$. It has recently been shown in [9] that Construction $A'$ is equivalent to the multilevel construction of Barnes-Wall lattices from Reed-Muller codes.

In this paper, we extend the lattice Constructions $D$, $D'$ and $\overline{D}$ from codes over $\mathbb{F}_p$ to linear codes over $\mathbb{Z}_q$, where $q \in \mathbb{N}$. We also provide a generalization of Construction $A'$ to codes over $\mathbb{Z}_q[X]/X^a$ and study the relationship between the proposed constructions. Section 2 is devoted to a brief summary of linear codes over the ring $\mathbb{Z}_q$. In Section 3.1, we present the extended Construction $D$ and show that it always produces a lattice. It is shown that a lattice obtained via extended Construction $D$ from a chain of $k$ nested linear codes over $\mathbb{Z}_q$ can, up to a scaling factor, always be obtained from Construction $A$ using a linear code over $\mathbb{Z}_{q^k}$. Furthermore, we show that various results presented in [1, 3] for lattices obtained via Construction $D$ from a family of nested binary linear codes also apply, with the necessary modifications, to lattices obtained from the extended Construction $D$ proposed here. The extended Constructions $D'$ and $\overline{D}$ are presented in sections 3.2 and 3.3, respectively. In Section 4, the relationship between Constructions $D$ and $\overline{D}$, is discussed and it is provided a necessary and sufficient condition for the set $\Gamma_{\overline{D}}$ obtained via Construction $\overline{D}$ to be a lattice. In Section 5, we provide a generalization to the real case of the recently developed Construction $A'$ and show that this construction produces a lattice if and only if the corresponding code over $\mathbb{Z}_q[X]/X^a$ is closed under a shifted zero-one addition. It is also established the relationship between Constructions $A'$ and $A$. Concluding remarks are included in Section 6.

## 2 Preliminaries

In this section, we present a brief introduction to linear codes over the ring $\mathbb{Z}_q$ of integers modulo $q$ ($q \in \mathbb{N}$), emphasizing some of their dissimilarities from codes over finite fields (see also [2,4,12] and their references).

A $q$-ary linear code $C$ of length $n$ over the ring $\mathbb{Z}_q$ is a $\mathbb{Z}_q$-submodule of $\mathbb{Z}_q^n$, that is, an additive subgroup of $\mathbb{Z}_q^n$. We denote by $C = \langle \boldsymbol{b}_1, ..., \boldsymbol{b}_k \rangle$ the code generated by the vectors $\boldsymbol{b}_1, ..., \boldsymbol{b}_k \in \mathbb{Z}_q^n$. A set $\{\boldsymbol{b}_1, ..., \boldsymbol{b}_k\}$ is a basis for $C$ when the vectors $\boldsymbol{b}_1, ..., \boldsymbol{b}_k$ are linearly independent and they generate $C$. In this case, any element $\boldsymbol{v} \in C$ can be written in a unique way as a linear combination of the vectors $\boldsymbol{b}_1, ..., \boldsymbol{b}_k$. In opposition to codes over fields there are $q$-ary linear codes that have no basis, for example, the code $C = \langle (2,4) \rangle = \{(0,0), (2,4), (4,8), (6,0), (8,4), (10,8)\} \subseteq \mathbb{Z}_{12}^2$ does not have basis because every nonempty subset of $C$ is linearly dependent ($6(a,b) = (0,0)$). Every $q$-ary linear code $C$ is characterized by a minimal set of generators [10, p. 64]. A generator matrix for a $q$-ary linear code $C$ is any matrix whose rows form a minimal set of generators for the code $C$.

We define an inner product on $\mathbb{Z}_q^n$ by $\boldsymbol{x} \cdot \boldsymbol{y} = x_1 y_1 + \cdots + x_n y_n$, where $\boldsymbol{x} = (x_1, ..., x_n)$, $\boldsymbol{y} = (y_1, ..., y_n) \in \mathbb{Z}_q^n$. Given a linear code $C \subseteq \mathbb{Z}_q^n$, the set $C^\perp = \{\boldsymbol{x} \in \mathbb{Z}_q^n;\ \boldsymbol{x} \cdot \boldsymbol{y} = 0,\ \forall \boldsymbol{y} \in C\}$ is also a linear code over $\mathbb{Z}_q$, which is called the dual code of $C$.

The following result is straightforward from the previous definitions. This result guarantees the existence of parameters to be chosen in Constructions $D$ and $D'$ in the next section.

**Theorem 1.** *Let $\mathbb{Z}_q^n \supseteq C_1 \supseteq C_2 \supseteq \cdots \supseteq C_a$ be a family of nested q-ary linear codes. Then, there are integers $k_1 \geq k_2 \geq \cdots \geq k_a \geq 0$ and $0 \leq r_1 \leq r_2 \leq \cdots \leq r_a$ and vectors $\boldsymbol{b}_1, ..., \boldsymbol{b}_{k_1}$, $\boldsymbol{h}_1, ..., \boldsymbol{h}_{r_a}$ in $\mathbb{Z}_q^n$ such that*

$$C_\ell = \left\langle \boldsymbol{b}_1, ..., \boldsymbol{b}_{k_\ell} \right\rangle \text{ and } C_\ell^\perp = \left\langle \boldsymbol{h}_1, ..., \boldsymbol{h}_{r_\ell} \right\rangle \text{ for all } \ell \in \{0, 1, ..., a\}.$$

*Moreover, if q is prime, the dimension of $C_i$ as a vector space over $\mathbb{Z}_q$ is $k_i$ and $r_i = n - k_i$ for all $i \in \{1, 2, ..., a\}$, then*
*$(i)$ there are vectors linearly independent $\boldsymbol{b}_1, ..., \boldsymbol{b}_{k_1}$ in $\mathbb{Z}_q^n$ such that*

$$C_\ell = \left\langle \boldsymbol{b}_1, ..., \boldsymbol{b}_{k_\ell} \right\rangle \text{ for all } \ell \in \{0, 1, ..., a\}.$$

*$(ii)$ there are linearly independent vectors $\boldsymbol{h}_1, ..., \boldsymbol{h}_{r_a}$ in $\mathbb{Z}_q^n$ such that*

$$C_\ell^\perp = \left\langle \boldsymbol{h}_1, ..., \boldsymbol{h}_{r_\ell} \right\rangle \text{ for all } \ell \in \{0, 1, ..., a\}.$$

We remark that when $q$ is not a prime number and $\mathbb{Z}_q^n \supseteq C_1 \supseteq C_2 \supseteq \cdots \supseteq C_a$ is a family of nested linear codes, there are not always integers $k_1 \geq k_2 \geq \cdots \geq k_a \geq 0$ and vectors $\boldsymbol{b}_1, ..., \boldsymbol{b}_{k_1}$ linearly independent in $\mathbb{Z}_q^n$ such that $\boldsymbol{b}_1, ..., \boldsymbol{b}_{k_\ell}$ span $C_\ell$, for $\ell = 1, ..., a$. For example, let $\mathbb{Z}_6^2 \supseteq C_1 \supseteq C_2$ be a family of nested linear codes, where $C_1 = \langle (1,2) \rangle$ and $C_2 = \langle (2,4) \rangle$. It is easy to see that there are no numbers $k_1 \geq k_2 \geq 0$ and vectors $\boldsymbol{b}_1, ..., \boldsymbol{b}_{k_1}$ linearly independent in $\mathbb{Z}_6^2$ such that $\boldsymbol{b}_1, ..., \boldsymbol{b}_{k_\ell}$ span $C_\ell$, for $\ell = 1, 2$.

## 3 Constructions $D$, $D'$ and $\overline{D}$ from codes over $\mathbb{Z}_q$

Let $\overline{\sigma}:\mathbb{Z}\to\mathbb{Z}_q$ be the natural ring homomorphism and $\sigma:\mathbb{Z}_q\to\mathbb{Z}$ be the standard inclusion map such that $\overline{\sigma}\big(\sigma(x)\big)=x$ for all $x\in\mathbb{Z}_q$. The map $\sigma$ is naturally extended to $\mathbb{Z}_q^n$ as $\sigma(x_1,...,x_n)=\big(\sigma(x_1),...,\sigma(x_n)\big)$.

### 3.1 Construction $D$

Next we introduce the notion of Construction $D$ from codes over $\mathbb{Z}_q$ and extend some results previously known for binary codes.

**Definition 1.** (*Construction D*) *Let $\mathbb{Z}_q^n\supseteq C_1\supseteq C_2\supseteq\cdots\supseteq C_a$ be a family of nested q-ary linear codes. Choose integers $k_1\geq k_2\geq\cdots\geq k_a\geq 0$ and vectors $\boldsymbol{b}_1,...,\boldsymbol{b}_{k_1}$ in $\mathbb{Z}_q^n$ such that $\boldsymbol{b}_1,...,\boldsymbol{b}_{k_\ell}$ span $C_\ell$, for $\ell=1,...,a$. The set $\Lambda_D$ consists of all vectors of the form*

$$\boldsymbol{z}+\sum_{\ell=1}^{a}\sum_{j=1}^{k_\ell}\beta_j^{(\ell)}\frac{1}{q^{\ell-1}}\sigma(\boldsymbol{b}_j),$$

*where $\boldsymbol{z}\in q\mathbb{Z}^n$ and $\beta_j^{(\ell)}\in\{0,1,...,q-1\}$.*

Note that to guarantee the existence of parameters $k_1\geq k_2\geq\cdots\geq k_a\geq 0$ and $\boldsymbol{b}_1,...,\boldsymbol{b}_{k_1}\in\mathbb{Z}_q^n$ in the definition above, we do not require that the vectors $\boldsymbol{b}_1,...,\boldsymbol{b}_{k_1}$ are linearly independent, since parameters satisfying this condition are only guaranteed when $q$ is prime (Theorem 1).

When $a=1$, Construction $D$ coincides with Construction $A$ for linear codes over $\mathbb{Z}_q$, $q\in\mathbb{N}$. Moreover, when $q=2$, each linear code $C_i$ can be seen as a vector subspace of $\mathbb{Z}_2^n$. Thus, if in Definition 1 we choose as parameters $q=2$, integers $k_i=\dim C_i$ $(i=1,...,a)$ and vectors linearly independent $\boldsymbol{b}_1,...,\boldsymbol{b}_{k_1}\in\mathbb{Z}_2^n$ such that some rearrangement of $\boldsymbol{b}_1,...,\boldsymbol{b}_{k_1}$ forms the rows of an "upper triangular" matrix, then (except for restrictions on the minimum distance of codes used) we obtain the classical Construction $D$ presented in [3,1].

The next theorem provides a new representation for the set $\Lambda_D$, which will be used in the following examples and theorems. In what follows, unless otherwise specified, $k_{a+1}:=0$.

**Theorem 2.**

$$\Lambda_D=\left\{\boldsymbol{z}+\sum_{s=1}^{a}\sum_{i=k_{s+1}+1}^{k_s}\alpha_i^{(s)}\frac{1}{q^{s-1}}\sigma(\boldsymbol{b}_i)\,\middle|\,\boldsymbol{z}\in q\mathbb{Z}^n,\alpha_i^{(s)}\in\mathbb{Z}\text{ and }0\leq\alpha_i^{(s)}<q^s\right\}.$$

*Proof.* Note that $k_1\geq k_2\geq\cdots\geq k_a\geq 0$, thus

$$\sum_{\ell=1}^{a}\sum_{j=1}^{k_\ell}\beta_j^{(\ell)}\frac{1}{q^{\ell-1}}\sigma(\boldsymbol{b}_j)=$$

$$=\left(\sum_{\ell=1}^{a}\beta_1^{(\ell)}q^{a-\ell}\right)\frac{1}{q^{a-1}}\sigma(\boldsymbol{b}_1)+\cdots+\left(\sum_{\ell=1}^{a}\beta_{k_a}^{(\ell)}q^{a-\ell}\right)\frac{1}{q^{a-1}}\sigma(\boldsymbol{b}_{k_a})+$$

$$\left(\sum_{\ell=1}^{a-1}\beta_{k_a+1}^{(\ell)}q^{a-1-\ell}\right)\frac{1}{q^{a-2}}\sigma(\boldsymbol{b}_{k_a+1})+\cdots+\left(\sum_{\ell=1}^{a-1}\beta_{k_{a-1}}^{(\ell)}q^{a-1-\ell}\right)\frac{1}{q^{a-2}}\sigma(\boldsymbol{b}_{k_{a-1}})$$

$$+\cdots+$$

$$\left(\beta_{k_2+1}^{(1)}\right)\frac{1}{q^0}\sigma(\boldsymbol{b}_{k_2+1})+\cdots+\left(\beta_{k_1}^{(1)}\right)\frac{1}{q^0}\sigma(\boldsymbol{b}_{k_1})$$

$$= \sum_{s=1}^{a} \sum_{i=k_{s+1}+1}^{k_s} \alpha_i^{(s)} \frac{1}{q^{s-1}} \sigma(\boldsymbol{b}_i), \text{ where } \alpha_i^{(s)} = \sum_{\ell=1}^{s} \beta_i^{(\ell)} q^{s-\ell} \text{ for all } 1 \le s \le a \quad (k_{a+1} := 0).$$

To complete this proof, is sufficient to observe that an integer $m$ satisfies $0 \le m < q^s$ if and only if there are $\beta_i^{(1)}, \dots, \beta_i^{(s)} \in \{0, 1, \dots, q-1\}$ such that $\sum_{\ell=1}^{s} \beta_i^{(\ell)} q^{s-\ell} = m$. □

**Theorem 3.** *$\Lambda_D$ is a full rank lattice.*

*Proof.* First note that $\Lambda_D$ is a discrete set, since $q^{a-1}\Lambda_D \subseteq \mathbb{Z}^n$. It remains to show that $\Lambda_D$ is an additive group. Indeed, $\Lambda_D$ is an additive group because $\mathbf{0} \in \Lambda_D$ and given $\boldsymbol{w}_1, \boldsymbol{w}_2 \in \Lambda_D$ we have by Theorem 2 that there are $\boldsymbol{z}_1, \boldsymbol{z}_2 \in q\mathbb{Z}^n$ and $0 \le \alpha_i^{(s)}, \beta_i^{(s)} < q^s$ such that

$$\boldsymbol{w}_1 = \boldsymbol{z}_1 + \sum_{s=1}^{a} \sum_{i=k_{s+1}+1}^{k_s} \alpha_i^{(s)} \frac{1}{q^{s-1}} \sigma(\boldsymbol{b}_i) \quad \text{and } \boldsymbol{w}_2 = \boldsymbol{z}_2 + \sum_{s=1}^{a} \sum_{i=k_{s+1}+1}^{k_s} \beta_i^{(s)} \frac{1}{q^{s-1}} \sigma(\boldsymbol{b}_i).$$

Then,

$$\boldsymbol{w}_1 - \boldsymbol{w}_2 = (\boldsymbol{z}_1 - \boldsymbol{z}_2) + \sum_{s=1}^{a} \sum_{i=k_{s+1}+1}^{k_s} \left(\alpha_i^{(s)} - \beta_i^{(s)}\right) \frac{1}{q^{s-1}} \sigma(\boldsymbol{b}_i).$$

Applying the division algorithm we obtain integers $\gamma_i^{(s)}$ and $\mu_i^{(s)}$, $1 \le s \le a$ and $k_{s+1} + 1 \le i \le k_s$, such that $\alpha_i^{(s)} - \beta_i^{(s)} = \gamma_i^{(s)} q^s + \mu_i^{(s)}$ and $0 \le \mu_i^{(s)} < q^s$. So we have

$$\boldsymbol{w}_1 - \boldsymbol{w}_2 = \boldsymbol{z} + \sum_{s=1}^{a} \sum_{i=k_{s+1}+1}^{k_s} \mu_i^{(s)} \frac{1}{q^{s-1}} \sigma(\boldsymbol{b}_i),$$

where

$$\boldsymbol{z} = \boldsymbol{z}_1 - \boldsymbol{z}_2 + q \sum_{s=1}^{a} \sum_{i=k_{s+1}+1}^{k_s} \gamma_i^{(s)} \sigma(\boldsymbol{b}_i) \in q\mathbb{Z}^n,$$

and by Theorem 2, we have $\boldsymbol{w}_1 - \boldsymbol{w}_2 \in \Lambda_D$. Therefore $\Lambda_D$ is a lattice. Moreover, $\Lambda_D$ has full rank because $q\mathbb{Z}^n \subseteq \Lambda_D$. □

**Theorem 4.** *If the vectors $\boldsymbol{b}_1, \dots, \boldsymbol{b}_{k_1}$ are linearly independent in $\mathbb{Z}_q^n$, then*

$$\#\left(\Lambda_D \cap [0, q)^n\right) = \prod_{s=1}^{a} \left(q^s\right)^{k_s - k_{s+1}}.$$

*Proof.* Note that since $\sigma(\boldsymbol{b}_1), \dots, \sigma(\boldsymbol{b}_{k_1})$ are vectors linearly independent in $\mathbb{R}^n$, there are exactly $\prod_{s=1}^{a} \left(q^s\right)^{k_s - k_{s+1}}$ vectors of the form

$$\sum_{s=1}^{a} \sum_{i=k_{s+1}+1}^{k_s} \alpha_i^{(s)} \frac{1}{q^{s-1}} \sigma(\boldsymbol{b}_i) \text{ with } 0 \le \alpha_i^{(s)} < q^s.$$

We have then to show that those elements are also distinct when considered modulo $q$. Thus, to complete the proof it is sufficient to observe that if

$$\boldsymbol{w}_1 = \sum_{s=1}^{a} \sum_{i=k_{s+1}+1}^{k_s} \alpha_i^{(s)} \frac{1}{q^{s-1}} \sigma(\boldsymbol{b}_i) \quad \text{and} \quad \boldsymbol{w}_2 = \sum_{s=1}^{a} \sum_{i=k_{s+1}+1}^{k_s} \beta_i^{(s)} \frac{1}{q^{s-1}} \sigma(\boldsymbol{b}_i),$$

where $0 \leq \alpha_i^{(s)} < q^s$ and $0 \leq \beta_i^{(s)} < q^s$ are congruent modulo $q$, then $\boldsymbol{w}_1 = \boldsymbol{w}_2$. Indeed, note that $\boldsymbol{w}_1 \equiv \boldsymbol{w}_2 \mod q$ if and only if

$$\sum_{s=1}^{a} \sum_{i=k_{s+1}+1}^{k_s} \left(\alpha_i^{(s)} - \beta_i^{(s)}\right) \frac{1}{q^{s-1}} \sigma(\boldsymbol{b}_i) \in q\mathbb{Z}^n.$$

Since the vectors $\boldsymbol{b}_1, \ldots, \boldsymbol{b}_{k_1}$ are linearly independent (over $\mathbb{Z}_q$), it follows that

$$\left(\alpha_i^{(s)} - \beta_i^{(s)}\right) \frac{1}{q^{s-1}} \in q\mathbb{Z}.$$

But $0 \leq \alpha_i^{(s)} < q^s$ and $0 \leq \beta_i^{(s)} < q^s$, ie, $-q < \left(\alpha_i^{(s)} - \beta_i^{(s)}\right) \frac{1}{q^{s-1}} < q$, thus

$$\left(\alpha_i^{(s)} - \beta_i^{(s)}\right) \frac{1}{q^{s-1}} = 0,$$

ie, $\alpha_i^{(s)} = \beta_i^{(s)}$, for every pair $(i,s)$ satisfying $1 \leq s \leq a$ and $k_{s+1} + 1 \leq i \leq k_s$. Therefore $\boldsymbol{w}_1 = \boldsymbol{w}_2$. □

In [5] it is shown that a lattice $\Lambda_D$ obtained via Construction $D$ from a chain of linear codes $\mathbb{Z}_q^n \supseteq C_1 \supseteq C_2 \supseteq \cdots \supseteq C_a$, where $q$ is prime, $k_i = \dim C_i$ $(i = 1, \ldots, a)$ and $\boldsymbol{b}_1, \ldots, \boldsymbol{b}_{k_1} \in \mathbb{Z}_q^n$ are vectors linearly independent, can also be obtained via Construction $A$ from a specific $q^a$-ary linear code. The next theorem extends this result to the generalized Construction $D$ from codes in $\mathbb{Z}_q^n$, $q \in \mathbb{N}$, considered here. We assume the notations of Definition 1.

**Theorem 5.** *Let $G_1$ be a matrix whose rows are the vectors $\sigma(\boldsymbol{b}_1), \ldots, \sigma(\boldsymbol{b}_{k_1})$ and let $C$ be the $q^a$-ary linear code generated by the rows of the matrix $G = DG_1$, where $D = (d_{ij})$ is a diagonal matrix such that*

$$d_{jj} = \begin{cases} 1, \text{ when } 1 \leq j \leq k_a \\ q, \text{ when } k_a < j \leq k_{a-1} \\ \vdots \\ q^{a-1}, \text{ when } k_2 < j \leq k_1. \end{cases}$$

*Therefore $q^{a-1}\Lambda_D = \Lambda_A(C)$.*

*Proof.* Let $\boldsymbol{w} \in \Lambda_A(C)$. Hence, it follows that there are $\alpha_j^{(i)} \in \{0, 1, \ldots, q^a - 1\}$, $1 \leq i \leq a$ and $k_{i+1} + 1 \leq j \leq k_i$, and a vector $\boldsymbol{z} \in \mathbb{Z}^n$ such that

$$\boldsymbol{w} = q^a \boldsymbol{z} + \sum_{i=1}^{a} \sum_{j=k_{i+1}+1}^{k_i} \alpha_j^{(i)} q^{a-i} \sigma(\boldsymbol{b}_j).$$

On the other hand, $q^{a-1}\Lambda_D$ is a lattice and $q^a\boldsymbol{z}, q^{a-i}\sigma(\boldsymbol{b}_j) \in q^{a-1}\Lambda_D$ $(1 \leq i \leq a$ and $k_{i+1} < j \leq k_i)$, and hence $\boldsymbol{w} \in q^{a-1}\Lambda_D$. This shows that $\Lambda_A(C) \subseteq q^{a-1}\Lambda_D$. To show that $q^{a-1}\Lambda_D \subseteq \Lambda_A(C)$, let $\boldsymbol{w} \in q^{a-1}\Lambda_D$. By Theorem 2, there are $\alpha_j^{(i)} \in \{0, 1, \ldots, q^i - 1\} \subseteq \{0, 1, \ldots, q^a - 1\}$, $1 \leq i \leq a$ and $k_{i+1} + 1 \leq j \leq k_i$, and a vector $\boldsymbol{z} \in \mathbb{Z}^n$ such that

$$\boldsymbol{w} = q^a \boldsymbol{z} + \sum_{i=1}^{a} \sum_{j=k_{i+1}+1}^{k_i} \alpha_j^{(i)} q^{a-i} \sigma(\boldsymbol{b}_j)$$

and consequently $\boldsymbol{w} \in \Lambda_A(C)$. □

The next theorem and Corollary 1 extend results previously known for binary codes. Corollary 2 can be found in [1,3].

**Theorem 6.** *Let $\boldsymbol{b}_1,...,\boldsymbol{b}_{k_1} \in \mathbb{Z}_q^n$ be nonzero vectors such that*

1. $C_\ell = \langle \boldsymbol{b}_1,...,\boldsymbol{b}_{k_\ell} \rangle$ *for* $\ell = 1,...,a$.
2. *Some row permutation of the matrix $M$, whose rows are $\boldsymbol{b}_1,...,\boldsymbol{b}_{k_1}$, forms an "upper triangular" (respectively, "lower triangular") matrix in the row echelon form.*
3. *The first nonzero component (respectively the last component) of each vector $\sigma(\boldsymbol{b}_i)$, $i = 1,...,k_1$, divides $q$ as well as all other components of this vector.*

*Then there is a basis for the lattice $\Lambda_D$ formed by $k_1$ vectors $(1/q^{i-1})\sigma(\boldsymbol{b}_j)$, where $1 \leq i \leq a$ and $k_{i+1} < j \leq k_i$, plus $n - k_1$ vectors of the shape $(0,...,0,q,0,...,0)$.*

*Proof.* Let us assume $M$ is an "upper triangular" matrix. Let $\tilde{M}$ be the order $n$ upper triangular matrix, whose rows are the $k_1$ vectors

$$\frac{1}{q^{i-1}}\sigma(\boldsymbol{b}_j), \text{ where } 1 \leq i \leq a \text{ and } k_{i+1} < j \leq k_i$$

and the others $n - k_1$ rows are of the form $(0,...,0,q,0,...,0)$. The existence and uniqueness of the matrix $\tilde{M}$ is guaranteed by the hypothesis 2, since $\boldsymbol{b}_1,...,\boldsymbol{b}_{k_1}$ are nonzero vectors. Note that the rows of the matrix $\tilde{M}$ are linearly independent, because if we denote by $\alpha_j$ $(1 \leq j \leq k_1)$ the first nonzero component of $\sigma(\boldsymbol{b}_j)$, we obtain

$$\det \tilde{M} = q^{n-k_1}\left(\prod_{j=1}^{k_1}\alpha_j\right)\prod_{i=1}^{a}\left(\frac{1}{q^{i-1}}\right)^{k_i-k_{i+1}} \neq 0.$$

Let $\Lambda(\tilde{M})$ be the lattice generated by the rows of the matrix $\tilde{M}$. To complete the proof, we will show that $\Lambda(\tilde{M}) = \Lambda_D$. In fact, let $\boldsymbol{w} \in \Lambda(\tilde{M})$. Hence, there are integers $\beta_j^{(i)}$, where $1 \leq i \leq a$ and $k_{i+1}+1 \leq j \leq k_i$ and a vector $\boldsymbol{z} \in q\mathbb{Z}^n$ such that

$$\boldsymbol{w} = \boldsymbol{z} + \sum_{i=1}^{a}\sum_{j=k_{i+1}+1}^{k_i}\beta_j^{(i)}\frac{1}{q^{i-1}}\sigma(\boldsymbol{b}_j).$$

Applying the division algorithm we obtain integers $\mu_j^{(i)}$ and $\alpha_j^{(i)}$, $1 \leq i \leq a$ and $k_{i+1}+1 \leq j \leq k_i$, such that $\beta_j^{(i)} = \mu_j^{(i)}q^i + \alpha_j^{(i)}$ and $0 \leq \alpha_j^{(i)} < q^i$. Thus,

$$\boldsymbol{w} = \tilde{\boldsymbol{z}} + \sum_{i=1}^{a}\sum_{j=k_{i+1}+1}^{k_i}\alpha_j^{(i)}\frac{1}{q^{i-1}}\sigma(\boldsymbol{b}_j), \text{ where } \tilde{\boldsymbol{z}} = \boldsymbol{z} + q\sum_{i=1}^{a}\sum_{j=k_{i+1}+1}^{k_i}\mu_j^{(i)}\sigma(\boldsymbol{b}_j) \in q\mathbb{Z}^n.$$

Therefore $\boldsymbol{w} \in \Lambda_D$ (Theorem 2). Reciprocally, if $\boldsymbol{w} \in \Lambda_D$, we have (from Theorem 2) that there are integers $\alpha_j^{(i)} \in \{0,1,...,q^i-1\}$, $1 \leq i \leq a$ and $k_{i+1}+1 \leq j \leq k_i$, and a vector $\boldsymbol{z} \in q\mathbb{Z}^n$ such that

$$\boldsymbol{w} = \boldsymbol{z} + \sum_{i=1}^{a}\sum_{j=k_{i+1}+1}^{k_i}\alpha_j^{(i)}\frac{1}{q^{i-1}}\sigma(\boldsymbol{b}_j).$$

Observe that

$$\sum_{i=1}^{a}\sum_{j=k_{i+1}+1}^{k_i}\alpha_j^{(i)}\frac{1}{q^{i-1}}\sigma(\boldsymbol{b}_j) \in \Lambda(\tilde{M}),$$

since the vectors $(1/q^{i-1})\sigma(\boldsymbol{b}_j)$, where $1 \leq i \leq a$ and $k_{i+1} < j \leq k_i$, are rows of the matrix $\tilde{M}$. Now, denote by $\boldsymbol{d}_1,...,\boldsymbol{d}_n$ the rows of the matrix $\tilde{M}$ and $\boldsymbol{e}_1,...,\boldsymbol{e}_n$ the canonical vectors of $\mathbb{R}^n$. Since the first nonzero component of each vector $\sigma(\boldsymbol{b}_i)$, $i = 1,...,k_1$, divides $q$ and all other components, each of the vectors $q\boldsymbol{e}_i$, $1 \leq i \leq n$, can be written as an integer linear combination of rows $\boldsymbol{d}_i, \boldsymbol{d}_{i+1},...,\boldsymbol{d}_n$. Thus, $q\mathbb{Z}^n \subseteq \Lambda(\tilde{M})$ and consequently, $\boldsymbol{z} \in \Lambda(\tilde{M})$. Hence, $\boldsymbol{w} \in \Lambda(\tilde{M})$. Therefore $\Lambda(\tilde{M}) = \Lambda_D$. □

We remark that, using the previous theorem, it is not difficult to see that (up to a scaling factor) some well-known lattices, such as $\mathbb{Z}^n, D_n, E_8, \Lambda_{16}$ and $\Lambda_{24}$ [3, Chapter 4] can be obtained via Construction $D$. For example, from the chain

$$\mathbb{Z}_8^{24} \supseteq \langle \boldsymbol{b}_1,...,\boldsymbol{b}_{24} \rangle \supseteq \langle \boldsymbol{b}_1,...,\boldsymbol{b}_{23} \rangle,$$

where $\boldsymbol{b}_1,...,\boldsymbol{b}_{24}$ are the rows of the matrix below, we obtain $\Lambda_D = \frac{\sqrt{8}}{8}\Lambda_{24}$.

$$\begin{pmatrix}
4&4&0&0&0&0&0&0&0&0&0&0&0&0&0&0&0&0&0&0&0&0&0&0\\
4&0&4&0&0&0&0&0&0&0&0&0&0&0&0&0&0&0&0&0&0&0&0&0\\
4&0&0&4&0&0&0&0&0&0&0&0&0&0&0&0&0&0&0&0&0&0&0&0\\
4&0&0&0&4&0&0&0&0&0&0&0&0&0&0&0&0&0&0&0&0&0&0&0\\
4&0&0&0&0&4&0&0&0&0&0&0&0&0&0&0&0&0&0&0&0&0&0&0\\
4&0&0&0&0&0&4&0&0&0&0&0&0&0&0&0&0&0&0&0&0&0&0&0\\
2&2&2&2&2&2&2&2&0&0&0&0&0&0&0&0&0&0&0&0&0&0&0&0\\
4&0&0&0&0&0&0&0&4&0&0&0&0&0&0&0&0&0&0&0&0&0&0&0\\
4&0&0&0&0&0&0&0&0&4&0&0&0&0&0&0&0&0&0&0&0&0&0&0\\
4&0&0&0&0&0&0&0&0&0&4&0&0&0&0&0&0&0&0&0&0&0&0&0\\
2&2&2&2&0&0&0&0&2&2&2&2&0&0&0&0&0&0&0&0&0&0&0&0\\
4&0&0&0&0&0&0&0&0&0&0&0&4&0&0&0&0&0&0&0&0&0&0&0\\
2&2&0&0&2&2&0&0&2&2&0&0&2&2&0&0&0&0&0&0&0&0&0&0\\
2&0&2&0&2&0&2&0&2&0&2&0&2&0&2&0&0&0&0&0&0&0&0&0\\
2&0&0&2&0&0&2&0&0&2&0&0&2&0&0&2&0&0&0&0&0&0&0&0\\
4&0&0&0&0&0&0&0&0&0&0&0&0&0&0&0&4&0&0&0&0&0&0&0\\
2&0&2&0&2&0&0&2&2&2&0&0&0&0&0&0&2&2&0&0&0&0&0&0\\
2&0&0&2&2&2&0&0&2&0&2&0&0&0&0&0&2&0&2&0&0&0&0&0\\
2&2&0&0&2&0&2&0&2&0&0&2&0&0&0&0&2&0&0&2&0&0&0&0\\
0&2&2&2&2&0&0&0&2&0&0&0&2&0&0&0&2&0&0&0&2&0&0&0\\
0&0&0&0&0&0&0&0&2&2&0&0&2&2&0&0&2&2&0&0&2&2&0&0\\
0&0&0&0&0&0&0&0&2&0&2&0&2&0&2&0&2&0&2&0&2&0&2&0\\
5&1&1&1&1&1&1&1&1&1&1&1&1&1&1&1&1&1&1&1&1&1&1&1\\
1&0&0&0&0&0&0&0&0&0&0&0&0&0&0&0&0&0&0&0&0&0&0&0
\end{pmatrix}$$

**Corollary 1.** *Let $\boldsymbol{b}_1,...,\boldsymbol{b}_{k_1} \in \mathbb{Z}_q^n$ be nonzero vectors satisfying the conditions* 1 *and* 2 *of Theorem 6. Suppose that the first nonzero component* (*respectively the last component*) *of each vector $\sigma(\boldsymbol{b}_i)$ is equal to* 1, *for $i = 1,...,k_1$. Then*

$$\det \Lambda_D = \left(q^2\right)^{n - \sum_{\ell=1}^{a} k_\ell}.$$

*Proof.* As we have seen in the previous theorem $\Lambda_D = \Lambda(\tilde{M})$, where $\tilde{M}$ is the upper triangular matrix of order $n$, whose rows are the $k_1$ vectors $(1/q^{i-1})\sigma(\boldsymbol{b}_j)$, where $1 \leq i \leq a$ and $k_{i+1} < j \leq k_i$, plus $n - k_1$ vectors of the shape $(0,...,0,q,0,...,0)$. Therefore

$$\det \Lambda_D = \det(\tilde{M}\tilde{M}^t) = \left(q^2\right)^{n - \sum_{\ell=1}^{a} k_\ell},$$

since

$$\begin{aligned}\det\tilde{M} &= q^{n-k_1}\prod_{i=1}^{a}\left(\frac{1}{q^{i-1}}\right)^{k_i-k_{i+1}} \\ &= q^{n-k_1}\cdot q^{-(k_a+k_{a-1}+\cdots+k_3+k_2)} \\ &= q^{n-\sum\limits_{\ell=1}^{a}k_\ell}.\end{aligned}$$

□

**Corollary 2.** *[1,3] Let $\boldsymbol{b}_1,...,\boldsymbol{b}_{k_1}\in\mathbb{Z}_2^n$ be nonzero vectors such that*

1. *$C_\ell=\langle \boldsymbol{b}_1,...,\boldsymbol{b}_{k_\ell}\rangle$ for $\ell=1,...,a$.*
2. *Some row permutation of the matrix $M$, whose rows are $\boldsymbol{b}_1,...,\boldsymbol{b}_{k_1}$, forms an "upper triangular" (respectively, "lower triangular") matrix in the row echelon form.*

*Then there is a basis for the lattice $\Lambda_D$ formed by $k_1$ vectors $(1/2^{i-1})\sigma(\boldsymbol{b}_j)$, where $1\le i\le a$ and $k_{i+1}<j\le k_i$, plus $n-k_1$ vectors of the shape $(0,...,0,2,0,...,0)$. Moreover,*

$$\det\Lambda_D = 4^{n-\sum\limits_{\ell=1}^{a}k_\ell}.$$

*Proof.* Note that for each $i\in\{1,...,k_1\}$ the first nonzero component of $\sigma(\boldsymbol{b_i})$ is equal to 1 and hence divides 2 as well as all other components of this vector. We obtain the desired result by applying Theorem 6 and Corollary 1. □

## 3.2 Construction $D'$

In this subsection we extend Construction $D'$ to codes over $\mathbb{Z}_q$, $q\in\mathbb{N}$.

**Definition 2.** (*Construction $D'$*) *Let $\mathbb{Z}_q^n\supseteq C_1\supseteq C_2\supseteq\cdots\supseteq C_a$ be a family of nested $q$-ary linear codes. Choose integers $r_1,r_2,...,r_a$ satisfying $0\le r_1\le r_2\le\cdots\le r_a$ and vectors $\boldsymbol{h}_1,...,\boldsymbol{h}_{r_a}$ in $\mathbb{Z}_q^n$ such that*

$$C_\ell^\perp=\langle \boldsymbol{h}_1,...,\boldsymbol{h}_{r_\ell}\rangle \text{ for } \ell=1,2,...,a$$

*where $C_\ell^\perp$ is the dual code of $C_\ell$. We define $\Lambda_{D'}$ as the set consisting of all vectors $\boldsymbol{x}\in\mathbb{Z}^n$ satisfying the congruences*

$$\boldsymbol{x}\cdot\sigma(\boldsymbol{h}_j)\equiv 0 \mod q^{i+1}$$

*for every pair $(i,j)$ satisfying $0\le i<a$ and $r_{a-i-1}<j\le r_{a-i}$, where $r_0:=0$.*

Note that to assure the existence of the parameters $0\le r_1\le r_2\le\cdots\le r_a$ and $\boldsymbol{h}_1,...,\boldsymbol{h}_{r_a}\in\mathbb{Z}_q^n$ in the above definition, it is not required that the vectors $\boldsymbol{h}_1,...,\boldsymbol{h}_{r_a}$ are linearly independent (the existence of parameters satisfying this condition is only guaranteed when $q$ is prime (Theorem 1)).

When $a=1$, Construction $D'$ coincides with the Construction $A$ for linear codes over $\mathbb{Z}_q$, $q\in\mathbb{N}$. Note also that if $q=2$, $r_i=n-\dim C_i$ $(i=1,...,a)$ and the vectors $\boldsymbol{h}_1,...,\boldsymbol{h}_{r_a}$ are linearly independent and such that some rearrangement forms the rows of an "upper triangular" matrix, then (except for restrictions on the minimum distance of the involved used codes) we obtain the classical Construction $D'$ presented in [1,3].

**Theorem 7.** *$\Lambda_{D'}$ is a full rank lattice.*

*Proof.* $\Lambda_{D'}$ is a discrete set, because $\Lambda_{D'} \subseteq \mathbb{Z}^n$. To see that $\Lambda_{D'}$ is an additive group note that $\mathbf{0} \in \Lambda_{D'}$ and given $\boldsymbol{x}, \boldsymbol{y} \in \Lambda_{D'}$ we have, for each pair of integers $(i, j)$ satisfying $0 \leq i < a$ and $r_{a-i-1} < j \leq r_{a-i}$,

$$\boldsymbol{x} \cdot \sigma(\boldsymbol{h_j}) \equiv 0 \mod q^{i+1} \quad \text{and} \quad \boldsymbol{y} \cdot \sigma(\boldsymbol{h_j}) \equiv 0 \mod q^{i+1}$$

and consequently, $(\boldsymbol{x}-\boldsymbol{y}) \cdot \sigma(\boldsymbol{h_j}) \equiv 0 \mod q^{i+1}$. Thus $\boldsymbol{x}-\boldsymbol{y} \in \Lambda_{D'}$. Therefore $\Lambda_{D'}$ is a lattice and $\Lambda_{D'}$ has full rank because $q^a\mathbb{Z}^n \subseteq \Lambda_{D'}$. □

### 3.3 Construction $\overline{D}$

Next we introduce Construction $\overline{D}$ associated with codes over $\mathbb{Z}_q$, $q \in \mathbb{N}$.

**Definition 3.** ($\textit{Construction } \overline{D}$) *Let $\mathbb{Z}_q^n \supseteq C_1 \supseteq C_2 \supseteq \cdots \supseteq C_a$ be a family of nested $q$-ary linear codes. We define the set $\Gamma_{\overline{D}}$ as follows*

$$\Gamma_{\overline{D}} := q^a\mathbb{Z}^n + q^{a-1}\sigma(C_1) + \cdots + q^{a-i}\sigma(C_i) + \cdots + q^1\sigma(C_{a-1}) + \sigma(C_a).$$

When $a = 1$, Construction $\overline{D}$ coincides with the Construction $A$ for linear codes over $\mathbb{Z}_q$. Moreover, when $q$ is prime, we obtain the Construction $\overline{D}$ from linear codes over $\mathbb{F}_q$.

Note that $\Gamma_{\overline{D}} \subseteq \mathbb{Z}^n$ is not always a lattice (as we will see in the Example 1). This remark leads us to the following definition.

**Definition 4.** *The lattice $\Lambda_{\overline{D}}$ is the smallest lattice that contains the set $\Gamma_{\overline{D}}$ (Definition 3). In other words,*

$$\Lambda_{\overline{D}} = \bigcap_{\Lambda \in \mathscr{L}} \Lambda$$

*where $\mathscr{L}$ is the set formed by all the lattices containing $\Gamma_{\overline{D}}$.*

In the next example we explore some dissimilarities between the Constructions $D$, $D'$ and $\overline{D}$.

**Example 1.** *Consider the nested 6-ary linear codes $\mathbb{Z}_6^2 \supseteq C_1 \supseteq C_2$, where*

$$\begin{aligned} C_1 &= \langle (1,2) \rangle \quad \textit{and} \\ C_2 &= \langle (2,4) \rangle . \end{aligned}$$

*Note that*

$$\begin{aligned} C_1^{\perp} &= \{(x,y) \in \mathbb{Z}_6^2; x+2y = 0\} \\ &= \{(0,0),(4,1),(2,2),(0,3),(4,4),(2,5)\} \\ &= \langle (4,1) \rangle . \end{aligned}$$

*and*

$$\begin{aligned} C_2^{\perp} &= \{(x,y) \in \mathbb{Z}_6^2; 2x+4y = 0\} \\ &= \{(0,0),(1,1),(4,1),(2,2),(5,2),(0,3), \\ &\quad (3,3),(1,4),(4,4),(2,5),(5,5),(3,0)\} \\ &= \langle (4,1),(3,0) \rangle . \end{aligned}$$

*Choosing the parameters* $k_1 = 2, k_2 = 1$, $r_1 = 1$, $r_2 = 2$ *and* $\boldsymbol{b}_1 = (2,4), \boldsymbol{b}_2 = (3,0), \boldsymbol{h}_1 = (4,1), \boldsymbol{h}_2 = (3,0) \in \mathbb{Z}_6^2$, *we have* $0 \leq k_2 \leq k_1$, $0 \leq r_1 \leq r_2$, $C_1 = \langle \boldsymbol{b}_1, \boldsymbol{b}_2 \rangle$, $C_2 = \langle \boldsymbol{b}_1 \rangle$, $C_1^{\perp} = \langle \boldsymbol{h}_1 \rangle$ *and* $C_2^{\perp} = \langle \boldsymbol{h}_1, \boldsymbol{h}_2 \rangle$. *Thus,*

$$\Lambda_D = \left\{ z + \alpha_2^{(1)}(3,0) + \alpha_1^{(2)} \frac{1}{6}(2,4) \,\middle|\, z \in 6\mathbb{Z}^2, 0 \leq \alpha_2^{(1)} \leq 5 \text{ and } 0 \leq \alpha_1^{(2)} \leq 35 \right\}$$

*and*

$$\Lambda_{D'} = \left\{ (x,y) \in \mathbb{Z}^2 \,\middle|\, 4x + y \equiv 0 \mod 36 \text{ and } 3x \equiv 0 \mod 6 \right\}.$$

*Also,*

$$\Gamma_{\overline{D}} = 6^2\mathbb{Z}^2 + 6^1\sigma(C_1) + 6^0\sigma(C_2),$$

*where*

$$\begin{aligned}
\sigma(C_1) &= \{(0,0),(1,2),(2,4),(3,0),(4,2),(5,4)\} \\
\sigma(C_2) &= \{(0,0),(2,4),(4,2)\}.
\end{aligned}$$

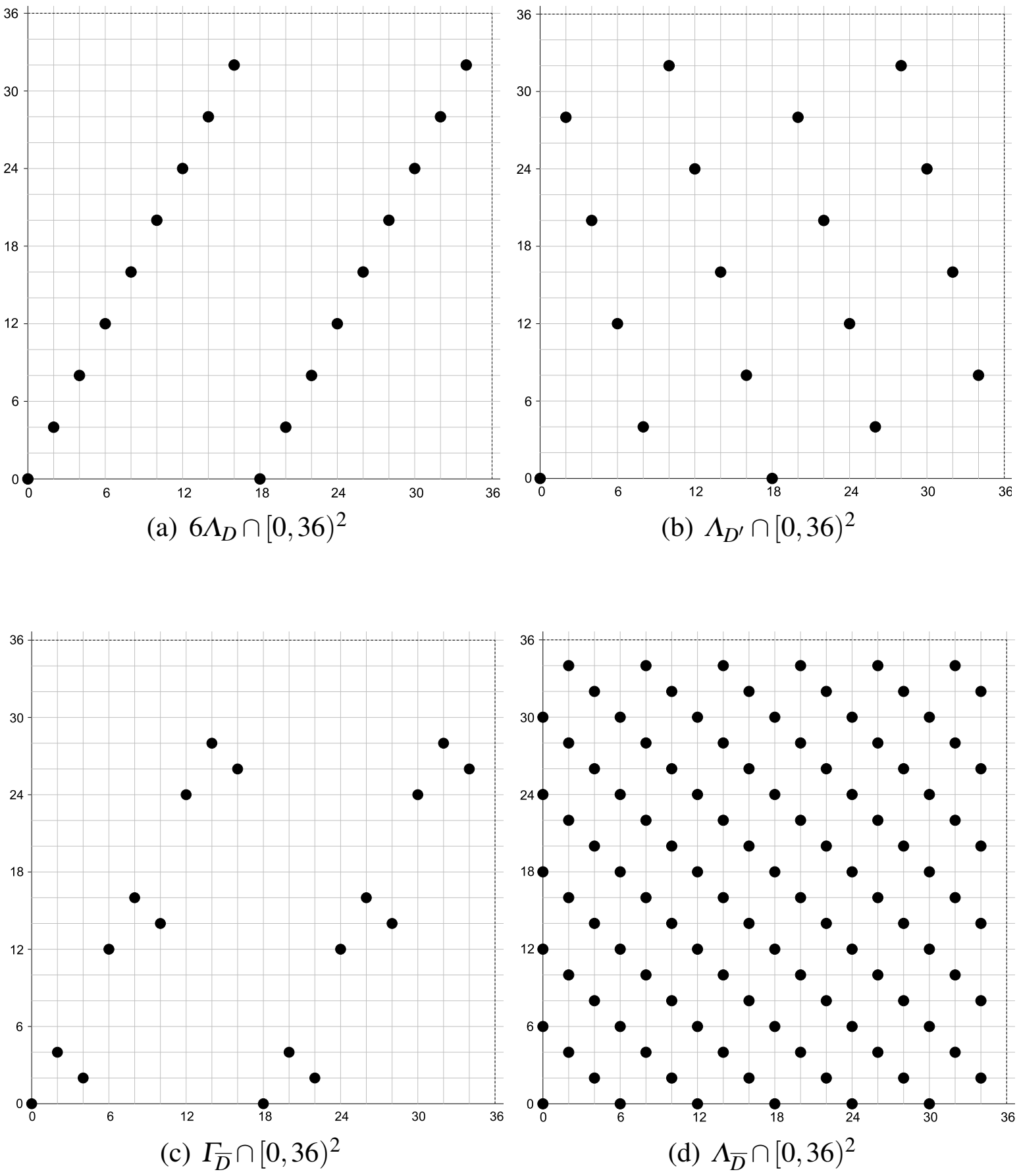

(a) $6\Lambda_D \cap [0,36)^2$ (b) $\Lambda_{D'} \cap [0,36)^2$

(c) $\Gamma_{\overline{D}} \cap [0,36)^2$ (d) $\Lambda_{\overline{D}} \cap [0,36)^2$

**Fig. 1** (Example 1) The elements of $6\Lambda_D$, $\Lambda_{D'}$, $\Gamma_{\overline{D}}$ and $\Lambda_{\overline{D}}$ inside the box $[0,36)^2$

*Therefore,*

$$\begin{aligned} 6\Lambda_D &= \bigcup_{z\in 36\mathbb{Z}^2} \left(z+(6\Lambda_D)\cap[0,36)^2\right), \\ \Lambda_{D'} &= \bigcup_{z\in 36\mathbb{Z}^2} \left(z+\Lambda_{D'}\cap[0,36)^2\right), \\ \Gamma_{\overline{D}} &= \bigcup_{z\in 36\mathbb{Z}^2} \left(z+\Gamma_{\overline{D}}\cap[0,36)^2\right) \text{ and} \\ \Lambda_{\overline{D}} &= \bigcup_{z\in 36\mathbb{Z}^2} \left(z+\Lambda_{\overline{D}}\cap[0,36)^2\right), \end{aligned}$$

*Note that* $(i)$ $\Gamma_{\overline{D}} \subsetneqq \Lambda_{\overline{D}}$, $(ii)$ $6\Lambda_D \subsetneqq \Lambda_{\overline{D}}$, $(iii)$ $6\Lambda_D \not\subset \Lambda_{D'}$, $(iv)$ $\Lambda_{D'} \not\subset 6\Lambda_D$ *and* $(v)$ $\Lambda_{D'} \subsetneqq \Lambda_{\overline{D}}$ *(see Figure 1).*

The next theorem and corollary of this section extend to lattices constructed from codes over $\mathbb{Z}_q$ results from [13] which deals with lattices from binary codes.

**Theorem 8.** *Let* $\mathbb{Z}_q^n \supseteq C_1 \supseteq C_2 \supseteq \cdots \supseteq C_a$ *be a family of nested q-ary linear codes. Then* $(i)$ $q^{a-1}\Lambda_D \subseteq \Lambda_{\overline{D}}$ *and* $(ii)$ $\Lambda_{\overline{D}}$ *consists of all vectors of the form*

$$q^a z + \sum_{i=1}^{a} q^{a-i} \sum_{c_j\in C_i} \alpha_j^{(i)} \sigma(c_j),$$

*where* $\alpha_j^{(i)} \in \{0,1,...,q-1\}$ *and* $z \in \mathbb{Z}^n$.

*Proof.* $(i)$ Let $k_1 \geq k_2 \geq \cdots \geq k_a \geq 0$ be integers and let $b_1,...,b_{k_1}$ be vectors in $\mathbb{Z}_q^n$ such that $C_i = \langle b_1,...,b_{k_i}\rangle$, for $i = 1,2,...,a$. Consider the lattice

$$\Lambda_D = \left\{ z + \sum_{i=1}^{a}\sum_{j=1}^{k_i} \alpha_j^{(i)} \frac{1}{q^{i-1}}\sigma(b_j) \,\middle|\, z\in q\mathbb{Z}^n \text{ and } \alpha_j^{(i)} \in \{0,1,...,q-1\} \right\}.$$

Let $w \in q^{a-1}\Lambda_D$,

$$w = q^a z + \sum_{i=1}^{a}\sum_{j=1}^{k_i} \alpha_j^{(i)} q^{a-i}\sigma(b_j).$$

Since $\Lambda_{\overline{D}}$ is a lattice, $q^a z \in q^a\mathbb{Z}^n \subseteq \Gamma_{\overline{D}} \subseteq \Lambda_{\overline{D}}$ and, for $1 \leq i \leq a$ and $1 \leq j \leq k_i$, we have $q^{a-i}\sigma(b_j) \in q^{a-i}\sigma(C_i) \subseteq \Gamma_{\overline{D}} \subseteq \Lambda_{\overline{D}}$. Therefore $w \in \Lambda_{\overline{D}}$.
$(ii)$ Let

$$\Lambda = \left\{ q^a z + \sum_{i=1}^{a} q^{a-i} \sum_{c_j\in C_i} \alpha_j^{(i)}\sigma(c_j) \,\middle|\, \alpha_j^{(i)} \in \{0,1,...,q-1\} \text{ and } z\in\mathbb{Z}^n \right\}.$$

Observe that $\Gamma_{\overline{D}} \subseteq \Lambda$, $\mathbf{0} \in \Lambda \subseteq \mathbb{Z}^n$ and

$$\sum_{i=1}^{a} q^{a-i} \sum_{c_j\in C_i} \alpha_j^{(i)}\sigma(c_j) = \sum_{i=1}^{a} q^{a-i} \sum_{c_j\in C_i\setminus C_{i+1}} \mu_j^{(i)}\sigma(c_j),$$

where $\mu_j^{(i)} = \alpha_j^{(i)} + \alpha_j^{(i-1)} q + \cdots + \alpha_j^{(1)} q^{i-1}$. Thus,

$$\Lambda = \left\{ q^a \boldsymbol{z} + \sum_{i=1}^{a} q^{a-i} \sum_{\boldsymbol{c}_j \in C_i \setminus C_{i+1}} \mu_j^{(i)} \sigma(\boldsymbol{c}_j) \,\middle|\, \mu_j^{(i)} \in \{0,1,...,q^i-1\} \text{ and } \boldsymbol{z} \in \mathbb{Z}^n \right\}.$$

Now, let $\boldsymbol{x}, \boldsymbol{y} \in \Lambda$. Then, there are $\lambda_j^{(i)}, \mu_j^{(i)} \in \{0,1,...,q^i-1\}$ and $\boldsymbol{z}_1, \boldsymbol{z}_2 \in \mathbb{Z}^n$ such that

$$\boldsymbol{x} = q^a \boldsymbol{z}_1 + \sum_{i=1}^{a} q^{a-i} \sum_{\boldsymbol{c}_j \in C_i \setminus C_{i+1}} \mu_j^{(i)} \sigma(c_j) \quad \text{and} \quad \boldsymbol{y} = q^a \boldsymbol{z}_2 + \sum_{i=1}^{a} q^{a-i} \sum_{\boldsymbol{c}_j \in C_i \setminus C_{i+1}} \lambda_j^{(i)} \sigma(\boldsymbol{c}_j).$$

Thus,

$$\boldsymbol{x} - \boldsymbol{y} = q^a (\boldsymbol{z}_1 - \boldsymbol{z}_2) + \sum_{i=1}^{a} q^{a-i} \sum_{\boldsymbol{c}_j \in C_i \setminus C_{i+1}} (\mu_j^{(i)} - \lambda_j^{(i)}) \sigma(\boldsymbol{c}_j).$$

Applying the division algorithm, we obtain integers $\beta_j^{(i)}$ and $r_j^{(i)}$ such that $0 \leq r_j^{(i)} < q^i$ and $\mu_j^{(i)} - \lambda_j^{(i)} = \beta_j^{(i)} q^i + r_j^{(i)}$ and so

$$\boldsymbol{x} - \boldsymbol{y} = q^a \boldsymbol{z} + \sum_{i=1}^{a} q^{a-i} \sum_{\boldsymbol{c}_j \in C_i \setminus C_{i+1}} r_j^{(i)} \sigma(\boldsymbol{c}_j), \text{ where } \boldsymbol{z} = \boldsymbol{z}_1 - \boldsymbol{z}_2 + \sum_{i=1}^{a} \sum_{\boldsymbol{c}_j \in C_i \setminus C_{i+1}} \beta_j^{(i)} \sigma(\boldsymbol{c}_j) \in \mathbb{Z}^n.$$

Thus $\boldsymbol{x} - \boldsymbol{y} \in \Lambda$ and hence $\Lambda$ is a lattice. To show that $\Lambda$ is the smallest lattice containing $\Gamma_{\overline{D}}$, let $\Lambda'$ be a lattice such that $\Gamma_{\overline{D}} \subseteq \Lambda'$ and let $\boldsymbol{v} \in \Lambda$. Hence, there are integers $\alpha_j^{(i)} \in \{0,1,...,q-1\}$ and a vector $\boldsymbol{z} \in \mathbb{Z}^n$ such that

$$\boldsymbol{v} = q^a \boldsymbol{z} + \sum_{i=1}^{a} q^{a-i} \sum_{\boldsymbol{c}_j \in C_i} \alpha_j^{(i)} \sigma(\boldsymbol{c}_j).$$

Now, since $q^a \boldsymbol{z} \in \Gamma_{\overline{D}}$ and $q^{a-i} \sigma(\boldsymbol{c}_j) \in q^{a-i} \sigma(C_i) \subseteq \Gamma_{\overline{D}}$ for all $i \in \{1,2,....,a\}$ and $\boldsymbol{c}_j \in C_i$, we have $\boldsymbol{v} \in \Lambda'$. Therefore $\Lambda$ is the smallest lattice $\Lambda_{\overline{D}}$ containing $\Gamma_{\overline{D}}$. □

**Corollary 3.** *Let $\mathscr{L}$ be the set of all lattices obtained from a family of nested q-ary linear codes $\mathbb{Z}_q^n = C_1 \supseteq C_2 \supseteq \cdots \supseteq C_a$ via Construction D. Then,*

$$\bigoplus_{\Lambda_D \in \mathscr{L}} (q^{a-1} \Lambda_D) = \Lambda_{\overline{D}}.$$

## 4 Connections between Constructions $D$ and $\overline{D}$

It has been shown in [13] that for $q = 2$ the Construction $\overline{D}$ produces a lattice if and only if the chain of nested linear codes is closed under the Schur product (known also as Hadamard product or componentwise multiplication). The next example shows that if $q \neq 2$ this result is not always true. Note also that not even when $q$ is prime we can guarantee this result.

**Example 2.** *Consider the nested* 5*-ary linear codes* $\mathbb{Z}_5^4 \supseteq C_1 \supseteq C_2$*, where*

$$\begin{aligned} C_1 &= \langle (1,2,3,4),(1,4,4,1)\rangle \ \text{ and} \\ C_2 &= \langle (1,2,3,4)\rangle . \end{aligned}$$

*Note that the chain* $\mathbb{Z}_5^4 \supseteq C_1 \supseteq C_2$ *is closed under the Schur product, but the set* $\Gamma_{\overline{D}}$ *obtained via Construction* $\overline{D}$ *from this chain is not a lattice. Indeed, note that*

$$C_2 = \{(0,0,0,0),(1,2,3,4),(2,4,1,3),(3,1,4,2),(4,3,2,1)\}$$

*and so*

$$\begin{aligned} \{\boldsymbol{x}\star\boldsymbol{y};\ \boldsymbol{x},\boldsymbol{y}\in C_2\} &= \{(0,0,0,0),(1,4,4,1),(2,3,3,2),(3,2,2,3),(4,1,1,4)\} \\ &= \langle (1,4,4,1)\rangle \subseteq C_1, \end{aligned}$$

*where* $\star$ *is the Schur product. Thus, the chain* $\mathbb{Z}_5^4 \supseteq C_1 \supseteq C_2$ *is closed under the Schur product. Now, note that* $\boldsymbol{v} = (1,2,3,4)$, $\boldsymbol{w} = (1,4,4,1) \in C_1$ *and* $\sigma(\boldsymbol{v}),\sigma(\boldsymbol{w}),\sigma(\boldsymbol{v}+\boldsymbol{w}) \in \Gamma_{\overline{D}}$*. But* $\sigma(\boldsymbol{v})+\sigma(\boldsymbol{w})-\sigma(\boldsymbol{v}+\boldsymbol{w}) = (0,5,5,5) \notin \Gamma_{\overline{D}}$ *and therefore* $\Gamma_{\overline{D}}$ *is not a lattice. Indeed, note that if we have* $\boldsymbol{c}_1 \in C_1, \boldsymbol{c}_2 \in C_2$ *and* $\boldsymbol{z} \in \mathbb{Z}^n$ *such that* $(0,5,5,5) = 25\boldsymbol{z} + 5\sigma(\boldsymbol{c}_1) + \sigma(\boldsymbol{c}_2)$*,* $\boldsymbol{c}_1 = (0,1,1,1)$*, but* $(0,1,1,1) \notin C_1$*.*

Taking into account the above remark we propose a new operation $*$ in $\mathbb{Z}_q^n$, which we call zero-one addition, as follows: For each pair of vectors $\boldsymbol{x} = (x_1,...,x_n)$ and $\boldsymbol{y} = (y_1,...,y_n)$ in $\mathbb{Z}_q^n$, the zero-one addition is given by

$$\boldsymbol{x} * \boldsymbol{y} := (x_1 * y_1, ..., x_n * y_n) \in \mathbb{Z}_q^n,$$

where

$$x_i * y_i = \begin{cases} 0, \text{ if } 0 \le \sigma(x_i)+\sigma(y_i) < q \\ 1, \text{ if } q \le \sigma(x_i)+\sigma(y_i) \le 2(q-1) \end{cases}$$

for $i = 1,...,n$. It is easy to see that

$$\sigma(\boldsymbol{x}) + \sigma(\boldsymbol{y}) = \sigma(\boldsymbol{x}+\boldsymbol{y}) + q\sigma(\boldsymbol{x}*\boldsymbol{y}). \tag{1}$$

Note that when $q = 2$ the zero-one addition coincides with the Schur product. We say that a family of nested $q$-ary linear codes $\mathbb{Z}_q^n \supseteq C_1 \supseteq C_2 \supseteq \cdots \supseteq C_a$ is closed under the zero-one addition if and only if the zero-one addition of any two elements of $C_i$ is contained in $C_{i-1}$, for $i = 2,...,a$. In other words, if $\boldsymbol{c}_1,\boldsymbol{c}_2 \in C_i$, then $\boldsymbol{c}_1 * \boldsymbol{c}_2 \in C_{i-1}$ for all $i = 2,...,a$.

**Example 3.** *Note that the chain of codes* $\mathbb{Z}_5^4 \supseteq C_1 \supseteq C_2$ *presented in the Example 2 is not closed under the zero-one addition. Indeed, suppose by contradiction that it is closed. Since* $(1,2,3,4)$, $(2,4,1,3)$, $(3,1,4,2) \in C_2$ *and*

$$\begin{aligned} (3,1,4,2)*(3,1,4,2) &= (1,0,1,0), \\ (2,4,1,3)*(2,4,1,3) &= (0,1,0,1), \\ (1,2,3,4)*(1,2,3,4) &= (0,0,1,1), \end{aligned}$$

*we conclude that* $(1,0,1,0),(0,1,0,1),(0,0,1,1) \in C_1$*. Therefore* $\dim C_1 \geq 3$ *and consequently,* $C_1 \neq \langle (1,2,3,4),(1,4,4,1)\rangle$ *(Contradiction).*

In the following theorem, we give a necessary and sufficient condition for the Construction $\overline{D}$ to produce a lattice. This theorem and its corollaries extend results presented in [13] for binary codes and their proofs can be carried out analogously.

**Theorem 9.** *Let $\mathbb{Z}_q^n \supseteq C_1 \supseteq C_2 \supseteq \cdots \supseteq C_a$ be a family of nested q-ary linear codes. The following statements are equivalent:*

1. *$\Gamma_{\overline{D}}$ is a lattice.*
2. *$\Gamma_{\overline{D}} = \Lambda_{\overline{D}}$.*
3. *$\mathbb{Z}_q^n \supseteq C_1 \supseteq C_2 \supseteq \cdots \supseteq C_a$ is closed under the zero-one addition.*
4. *$\Gamma_{\overline{D}} = q^{a-1}\Lambda_D$.*

As an immediate consequence of the previous theorem, we have the following result: If a family of nested $q$-ary linear codes $\mathbb{Z}_q^n \supseteq C_1 \supseteq C_2 \supseteq \cdots \supseteq C_a$ is closed under the zero-one addition, then Construction $\overline{D}$ and Construction $D$, except for a scaling factor, yield the same lattice (which does not depend on the parameters $k_1, k_2, \cdots, k_a$ or $\boldsymbol{b}_1, ..., \boldsymbol{b}_{k_1}$).

In the next example, we display a chain of linear codes that is closed under the zero-one addition, but is not closed under the Schur product and present its associated lattice.

**Example 4.** *Consider the nested 5-ary linear codes $\mathbb{Z}_5^4 \supseteq C_1 \supseteq C_2$, where*

$$\begin{aligned} C_1 &= \langle (1,0,1,0),(0,1,0,1),(0,0,1,1)\rangle \ \textit{and} \\ C_2 &= \langle (1,2,3,4)\rangle. \end{aligned}$$

*The chain $\mathbb{Z}_5^4 \supseteq C_1 \supseteq C_2$ is closed under the zero-one addition, but it is not closed under the Schur product. Indeed, note that*

$$C_2 = \{(0,0,0,0),(1,2,3,4),(2,4,1,3),(3,1,4,2),(4,3,2,1)\}$$

*and so*

$$\begin{aligned} \left\{\boldsymbol{x} * \boldsymbol{y};\ \boldsymbol{x},\boldsymbol{y} \in C_2\right\} &= \{(0,0,0,0),(1,0,1,0),(0,1,0,1), \\ &\quad (0,0,1,1),(1,1,1,1),(1,1,0,0)\} \subseteq C_1. \end{aligned}$$

*Therefore $\mathbb{Z}_5^4 \supseteq C_1 \supseteq C_2$ is closed the under zero-one addition.*

*Now, notice that $(1,2,3,4) \in C_2$ and $(1,2,3,4) \star (1,2,3,4) = (1,4,4,1) \notin C_1$.Therefore $\mathbb{Z}_5^4 \supseteq C_1 \supseteq C_2$ is not closed under the Schur product.*

*As pointed out in Theorem 9 the set $\Gamma_{\overline{D}}$ of this chain is a lattice and $\Gamma_{\overline{D}} = 5\Lambda_D$. So it is easy to see that*

$$\{(1,2,3,4),(0,5,0,5),(0,0,5,5),(0,0,0,25)\}$$

*is a basis for $\Gamma_{\overline{D}}$ (Theorem 6).*

## 5 Construction $A'$

We consider the polynomial quotient ring $R_{a,q} := \mathbb{Z}_q[X]/X^a$ where $X$ is a variable. A linear code over $R_{a,q}$ is a submodule of $R_{a,q}^n$. The code $C$ corresponding to a generator matrix $G \in R_{a,q}^{k\times n}$ is given by

$$C = \{\boldsymbol{u}G \mid \boldsymbol{u} \in R_{a,q}^{1\times k}\},$$

where the matrix multiplication is over the ring $R_{a,q}$. We define the mapping $\phi : R_{a,q} \to \mathbb{Z}$ as follows

$$\phi\left(\sum_{j=0}^{a-1} b_j X^j\right) = \sum_{j=0}^{a-1} \sigma(b_j) q^j,$$

and consider its extension from $R_{a,q}^n$ to $\mathbb{Z}^n$ as $\phi(p_1,...,p_n) = \big(\phi(p_1),...,\phi(p_n)\big)$.

Next, we introduce the Construction $A'$ associated to linear codes over $R_{a,q}$, which is a generalization of the real Construction $A'$ proposed by Harshan et al. in [8,9].

**Definition 5.** *(Construction $A'$) Given a linear code $C$ over $R_{a,q}$, we define the set*

$$\Gamma_{A'} = \phi(C) + q^a\mathbb{Z}^n.$$

*We say that the lattice obtained from Construction $A'$ using code $C$, denoted by $\Lambda_{A'}$, is the smallest lattice that contains $\Gamma_{A'}$.*

The next theorem shows that given a chain of $q$-ary linear codes, the set $\Gamma_{\overline{D}}$ obtained from Construction $\overline{D}$ can also be obtained from Construction $A'$, that is, there is always a linear code $C$ over $R_{a,q}$ such that $\Gamma_{\overline{D}} = \phi(C) + q^a\mathbb{Z}^n$. This result was presented in [13] for the case $q = 2$. The proof of this extended version can be carried out similarly by considering matrices whose rows generate the lattice (not requiring independent rows) instead of generator matrices, and so it will be omitted.

**Theorem 10.** *Let $\mathbb{Z}_q^n \supseteq C_1 \supseteq \cdots \supseteq C_{a-1} \supseteq C_a$ be a family of nested $q$-ary linear codes and*

$$\Gamma_{\overline{D}} := q^a\mathbb{Z}^n + q^{a-1}\sigma(C_1) + \cdots + q^{a-i}\sigma(C_i) + \cdots + q^1\sigma(C_{a-1}) + \sigma(C_a).$$

*There exists a linear code $C$ over $R_{a,q}$ such that $\Gamma_{\overline{D}} = \phi(C) + q^a\mathbb{Z}^n = \Gamma_{A'}$.*

**Corollary 4.** *If a linear code $C$ over $R_{a,q}$ can be written as*

$$C_a + XC_{a-1} + \cdots + X^{a-1}C_1,$$

*where $\mathbb{Z}_q^n \supseteq C_1 \supseteq \cdots \supseteq C_{a-1} \supseteq C_a$ is closed under the zero-one addition, then the set $\Gamma_{A'} = \phi(C) + q^a\mathbb{Z}^n$ obtained from Construction $A'$ is a lattice.*

*Proof.* Note that

$$\Gamma_{A'} = \phi(C) + q^a\mathbb{Z}^n = \sigma(C_a) + q\sigma(C_{a-1}) + \cdots + q^{a-1}\sigma(C_1) + q^a\mathbb{Z}^n = \Gamma_{\overline{D}}.$$

Theorem 9 concludes this proof. □

Let $p(X) = \alpha_0 + \alpha_1 X + \cdots + \alpha_{a-1}X^{a-1}$ and $\hat{p}(X) = \hat{\alpha}_0 + \hat{\alpha}_1 X + \cdots + \hat{\alpha}_{a-1}X^{a-1}$ be elements in $R_{a,q}$. We define zero-one addition of $p(X)$ and $\hat{p}(X)$ as

$$p(X) * \hat{p}(X) = \alpha_0 * \hat{\alpha}_0 + \alpha_1 * \hat{\alpha}_1 X + \cdots + \alpha_{a-1} * \hat{\alpha}_{a-1}X^{a-1} \in R_{a,q},$$

where the operation $*$ the right-hand side of the above equality is the zero-one addition in $\mathbb{Z}_q$. Now, for $\boldsymbol{p} = (p_1, ..., p_n)$ and $\hat{\boldsymbol{p}} = (\hat{p}_1, ..., \hat{p}_n)$ in $R_{a,q}^n$, we define

$$\boldsymbol{p} * \hat{\boldsymbol{p}} = (p_1 * \hat{p}_1, ..., p_n * \hat{p}_n).$$

Alternatively, if write $\boldsymbol{p}, \hat{\boldsymbol{p}} \in R_{a,q}^n$ as $\boldsymbol{p} = \boldsymbol{p}_0 + \boldsymbol{p}_1 X + \cdots + \boldsymbol{p}_{a-1}X^{a-1}$ and $\hat{\boldsymbol{p}} = \hat{\boldsymbol{p}}_0 + \hat{\boldsymbol{p}}_1 X + \cdots + \hat{\boldsymbol{p}}_{a-1}X^{a-1}$, where $\boldsymbol{p}_i, \hat{\boldsymbol{p}}_i \in \mathbb{Z}_q^n$ for $i = 0, ..., a-1$,

$$\boldsymbol{p} * \hat{\boldsymbol{p}} = (\boldsymbol{p}_0 * \hat{\boldsymbol{p}}_0) + (\boldsymbol{p}_1 * \hat{\boldsymbol{p}}_1)X + \cdots + (\boldsymbol{p}_{a-1} * \hat{\boldsymbol{p}}_{a-1})X^{a-1}.$$

We say that a linear code $C$ over $R_{a,q}$ is closed under the shifted zero-one addition if and only if, for any elements $\boldsymbol{c}_1$ and $\boldsymbol{c}_2$ of $C$, $(\boldsymbol{c}_1 * \boldsymbol{c}_2)X$ is also an element of $C$. In other words,

$$\boldsymbol{c}_1, \boldsymbol{c}_2 \in C \Longrightarrow (\boldsymbol{c}_1 * \boldsymbol{c}_2)X \in C.$$

We remark that, if a chain of $q$-ary linear codes $\mathbb{Z}_q^n \supseteq C_1 \supseteq \cdots \supseteq C_{a-1} \supseteq C_a$ is closed under the zero-one addition, then $C = C_a + XC_{a-1} + \cdots + X^{a-1}C_1$ is a linear code over $R_{a,q}^n$ closed under the shifted zero-one addition.

The next theorem provides a necessary and sufficient condition for Construction $A'$ to produce a lattice. This result was also stated and proved in [13] for the case $q = 2$ and the proof for the general case can be done analogously.

**Theorem 11.** *Let $C$ be a linear code over $R_{a,q}$. The set $\Gamma_{A'} = \phi(C) + q^a\mathbb{Z}^n$ obtained from Construction $A'$ is a lattice if and only if $C$ is closed under the shifted zero-one addition.*

As a consequence of above theorem we also can obtain the following result.

**Corollary 5.** *Let $\overline{\sigma}_{q^a} : \mathbb{Z}^n \to \mathbb{Z}^n_{q^a}$ be the canonical ring homomorphism. A linear code $C$ over $R_{a,q}$ is closed under the shifted zero-one addition if and only if $\overline{\sigma}_{q^a}(\phi(C))$ is a $q^a$-ary linear code. In this case,*

$$\Gamma_{A'} = \Lambda_A\big(\overline{\sigma}_{q^a}(\phi(C))\big).$$

*Proof.* ($\Rightarrow$) Note that $\overline{\sigma}_{q^a}(\phi(C)) = \overline{\sigma}_{q^a}(\Gamma_{A'})$ and $\Gamma_{A'}$ is a lattice, since $C$ is a linear code over $R_{a,q}$ and is closed under the shifted zero-one addition. Now, notice that $\overline{\sigma}_{q^a}(\phi(C)) \subseteq \mathbb{Z}^n_{q^a}$, $\mathbf{0} \in \overline{\sigma}_{q^a}(\phi(C))$ and given $\boldsymbol{w}_1, \boldsymbol{w}_2 \in \overline{\sigma}_{q^a}(\phi(C)) = \overline{\sigma}_{q^a}(\Gamma_{A'})$, there are $\boldsymbol{c}_1, \boldsymbol{c}_2 \in \Gamma_{A'}$ such that $\boldsymbol{w}_1 = \overline{\sigma}_{q^a}(\phi(\boldsymbol{c}_1))$ and $\boldsymbol{w}_2 = \overline{\sigma}_{q^a}(\phi(\boldsymbol{c}_2))$ and consequently

$$\begin{aligned}\boldsymbol{w}_1 - \boldsymbol{w}_2 &= \overline{\sigma}_{q^a}(\phi(\boldsymbol{c}_1)) - \overline{\sigma}_{q^a}(\phi(c_2)) \\ &= \overline{\sigma}_{q^a}\big(\phi(\boldsymbol{c}_1) - \phi(\boldsymbol{c}_2)\big) \in \overline{\sigma}_{q^a}(\phi(C)).\end{aligned}$$

($\Leftarrow$) Suppose that $C$ is not closed under the shifted zero-one addition. By Theorem 11, it follows that $\Gamma_{A'}$ is not a lattice. Thus, there are $\boldsymbol{v}_1, \boldsymbol{v}_2 \in \Gamma_{A'}$ such that $\boldsymbol{v}_1 - \boldsymbol{v}_2 \notin \Gamma_{A'}$, because $\Gamma_{A'}$ is a discrete set. Since $\boldsymbol{v}_1, \boldsymbol{v}_2 \in \Gamma_{A'}$, there are $\boldsymbol{c}_1, \boldsymbol{c}_2 \in C$ and $\boldsymbol{z}_1, \boldsymbol{z}_2 \in \mathbb{Z}^n$ such that $\boldsymbol{v}_1 = \phi(\boldsymbol{c}_1) + q^a\boldsymbol{z}_1$ and $v_2 = \phi(\boldsymbol{c}_2) + q^a\boldsymbol{z}_2$. Let $\boldsymbol{w}_1 = \overline{\sigma}_{q^a}(\boldsymbol{v}_1)$ and $\boldsymbol{w}_2 = \overline{\sigma}_{q^a}(\boldsymbol{v}_2)$,

$$\boldsymbol{w}_1 - \boldsymbol{w}_2 = \overline{\sigma}_{q^a}(\boldsymbol{v}_1) - \overline{\sigma}_{q^a}(\boldsymbol{v}_2) = \overline{\sigma}_{q^a}(\boldsymbol{v}_1 - \boldsymbol{v}_2) \notin \overline{\sigma}_{q^a}(\phi(C)),$$

because if $\overline{\sigma}_{q^a}(\boldsymbol{v}_1 - \boldsymbol{v}_2) = \overline{\sigma}_{q^a}(\phi(\boldsymbol{c}))$ for some $\boldsymbol{c} \in C$, then $\boldsymbol{v}_1 - \boldsymbol{v}_2 = \phi(\boldsymbol{c}) + q^a\boldsymbol{z}$ for some $\boldsymbol{z} \in \mathbb{Z}^n$, ie, $\boldsymbol{v}_1 - \boldsymbol{v}_2 \in \Gamma_{A'}$. Thus, $\overline{\sigma}_{q^a}(\phi(C))$ is not a $q^a$-ary linear code.

Therefore, a linear code $C$ over $R_{a,q}$ is closed under the shifted zero-one addition if and only if $\overline{\sigma}_{q^a}(\phi(C))$ is a $q^a$-ary linear code. Finally, note that if $\overline{\sigma}_{q^a}(\phi(C))$ is a $q^a$-ary linear code, then the lattice $q^a$-ary obtained from Construction $A$ using code $\overline{\sigma}_{q^a}(\phi(C))$ is

$$\Lambda_A\big(\overline{\sigma}_{q^a}(\phi(C))\big) = \sigma_{q^a}\big(\overline{\sigma}_{q^a}(\phi(C))\big) + q^a\mathbb{Z}^n = \phi(C) + q^a\mathbb{Z}^n = \Gamma_{A'}.$$

□

## 6 Conclusion

Constructions $D$, $D'$, $\overline{D}$ and $A'$ were extended here to codes over $\mathbb{Z}_q$. Constructions $D$ and $D'$ are always lattices. Construction $\overline{D}$ from linear codes over $\mathbb{Z}_q$ produces a lattice if and only if the nested codes being used are closed under the zero-one addition as derived here. Construction $A'$ from a linear code $C$ over $\mathbb{Z}_q[X]/X^a$ produces a lattice if and only if $C$ is closed under a shifted zero-one addition. Further interesting directions of research may include the study of minimum norm, kissing number and center density of the lattices obtained via Constructions $D$, $D'$ and $\overline{D}$ from linear codes over $\mathbb{Z}_q$, using different metrics such as the Lee metric. Chains of recently approached codes over $\mathbb{Z}_q$ [2,4,12] may be considered in the search for lattices with good properties.